\documentclass[11pt]{article}

\usepackage{graphicx}        
\usepackage[bottom]{footmisc}

\usepackage{newtxtext}       %
\usepackage{newtxmath}       

\usepackage{bbm}
\usepackage{amsmath}
\usepackage{hyperref}

\numberwithin{equation}{section}

\DeclareMathOperator*{\argmax}{arg\,max}

\usepackage[shortlabels]{enumitem}

\usepackage{caption}
\usepackage{subcaption}
\usepackage{graphicx}
\usepackage{multirow}

\makeindex             

\begin{document}

\title{A comparison of different clustering approaches for high-dimensional presence-absence data}
%
\author{Gabriele d'Angella\footnote{Università di Bologna, Dipartimento di Scienze Statistiche, gabriele.dangella@studio.unibo.it} and Christian Hennig\footnote{Università di Bologna, Dipartimento di Scienze Statistiche, christian.hennig@unibo.it}}

%
%
\maketitle


\abstract{Presence-absence data is defined by vectors or matrices of zeroes and ones, where the ones usually indicate a ``presence'' in a certain place. Presence-absence data occur for example when investigating geographical species distributions, genetic information, or the occurrence of certain terms in texts. There are many applications for clustering such data; one example is to find so-called biotic elements, i.e., groups of species that tend to occur together geographically. Presence-absence data can be clustered in various ways, namely using a latent class mixture approach with local independence, distance-based hierarchical clustering with the Jaccard distance, $K$-modes, a density-based approach, or also using clustering methods for continuous data on a multidimensional scaling representation of the distances. These methods are conceptually very different and can therefore not easily be compared theoretically. We compare their performance with a comprehensive simulation study based on models for species distributions.~\\~\\
{\it This has been accepted for publication in Ferreira, J., Bekker, A., Arashi, M. and Chen, D. (eds.) Innovations in multivariate statistical modelling: navigating theoretical and multidisciplinary domains, Springer Emerging Topics in Statistics and Biostatistics.}}    
~

\section{Introduction}
\label{sec1}
Presence-absence data comprises observations that are vectors or matrices of zeroes and ones, where the ones usually indicate a ``presence'' in a certain place. Presence-absence data are often high-dimensional (the number of ``places'' may be large compared to the number of observations), and occur in a large range of applications, for example when investigating geographical species distributions, genetic information, or the occurrence of certain terms in texts.  

Often there is an interest in clustering presence-absence data. An example for this is the search for ``biotic elements'', which are groups of species showing very similar distribution areas that can constitute evidence for the existence of areas of endemism generated by the formation of barriers over geologic time periods \cite{haus02,haus03}.

Here we present a simulation study for comparing different approaches to clustering such data. This is based on a model for generating artificial but realistic presence-absence data with known clusters, the recovery of which can be compared. Not only are we interested in ranking methods, we also investigate characteristics of the data (such as degree of cluster overlap) that drive the methods' performances; the comparison of methods may strongly depend on such characteristics. We will also explain in detail why some methods do not perform that well. See \cite{vanmech} for guidelines on benchmark studies to compare clustering methods.

There are various approaches to clustering presence-absence data. Two major approaches are (1) the use of distance-based clustering methods such as average linkage or partitioning around medoids \cite{everitt} using distance measures between the observations \cite{hazel,shi}, and (2) latent class mixture models \cite{hagenaars}. 
Some of the most popular clustering methods such as K-Means \cite{jain} and model-based clustering based on Gaussian mixtures \cite{mclust} require Euclidean data and cannot directly be applied to presence-absence data. As approach (3), Hausdorf and Hennig \cite{haus03,hennig04} have proposed to use methods for Euclidean data on the output of a Multidimensional Scaling (MDS) \cite{borg2005} specifically with presence-absence data for biotic element analysis. Furthermore, \cite{AzzMen16} generalised their density-based clustering in this way to categorical data.

MDS is a set of techniques that generate a Euclidean representation so that the resulting Euclidean distances approximate a given usually non-Euclidean distance structure. Low dimensionality of the representation is often desirable. Such a representation is often used for visualization (for which low dimensionality is needed), but can also be used for making distance data accessible to more elaborate statistical methodology that requires Euclidean input, as is done here for cluster analysis.

A disadvantage of this approach is that the MDS generally will lose some information in the original data (the lower the dimension, the larger the loss). A key interest in the present study is whether this approach can compete with the more direct approaches (1), and (2).

The relationship between clustering and scaling techniques has also been discussed in de Leeuw and Heiser \cite{deleeuw} and in Kruskal \cite{krus77}. Simultaneous use of MDS and clustering is treated by Desarbo \cite{desarbo} and Oh and Raftery \cite{msoh}. Some biogeographical work related to our study has been done by Vavrik \cite{vavrek} (comparison of clustering methods for fossil data) and Ulrich and Gotelli \cite{ulrich} (null models for simulating presence-absence data).   

Section \ref{sec2} has a basic description of the data, distance, and MDS. Section \ref{secclus} introduces the involved clustering methods. Section \ref{sec3} describes in detail how the data for the simulations were generated. Section \ref{sec4} discusses the results of the study, and Section \ref{sec5} concludes the paper. 

\section{Data and preprocessing}
\label{sec2}
Data here are $m$-dimensional binary: $\mathbf{x}_i=(x_{i1},\ldots,x_{im}),\ i=1,{\dots},n$, for all $j:\ x_{ij}\in\{0,1\}$. The data are modelled with a specific application area in mind, which is analysing presence (1) and absence (0) of species in geographical regions, where $i$ indexes the species and $j$ indexes the region. Clusters are called ``biotic elements'' and are of interest because they provide insight into natural history \cite{haus02,hennig04}. Our results may however also be informative for presence-absence data in other applications. 

Methods are compared regarding their capability to retrieve true clusters in the data as set in the simulation. The cluster membership of the observations is denoted as $\mathbf{c} = (c_1, \ldots, c_n)$, where $c_i=k$ denotes that observation $i$ is in cluster $k\in\{1,\ldots,K\}$.

Some of the compared clustering methods operate on distances. As distance we use the popular Jaccard distance \cite{jaccard,hazel,dws98}, for discussion and alternatives see \cite{hazel,shi,everitt,hennig06}:
\begin{equation}
d_J(\mathbf{x}_1,\mathbf{x}_2) = 1 - \frac{\sum_{j=1}^{m}  \mathbbm{1}(x_{1j}=1 \land x_{2j}=1)}{\sum_{j=1}^{m}  \mathbbm{1}(x_{1j}=1 \lor x_{2j}=1)}
\end{equation}
where $\mathbbm{1}$ denotes the indicator function.

A topic of key interest here is the performance of clustering techniques for Euclidean data that use an MDS representation of the distances as input. In this study, two MDS methods were applied: classical scaling \cite{torgerson} as implemented in the R-function \texttt{cmdscale}, and ratio MDS \cite{borg2005,borg2013} as implemented in the R-package \texttt{smacof}, function \texttt{mds}. Two and three-dimensional MDS solutions were used in the simulation study. Although this seems low, the results (as provided in Section \ref{sec4}) were good enough that there is not much room for improvement by higher dimensional solutions (in \cite{msoh}, lower-dimensional proxies yielded clustering solutions almost as good as those exploiting higher-dimensional configurations). An advantage of using a low dimensional MDS is the ease of visualization of the data. However, in more general applications, using a higher dimensional MDS solution may be helpful in case that too much information is lost by a low dimensional solution.

However, there are no indications against the setting of $p>3$ when it comes to the application of techniques like KM or GMM and this should be considered in case the comparison with other clustering recovery methods shows that too small a $p$ might lose too much information. In this project we deal with multivariate binary data: not only does MDS constitute a powerful visualization tool, it also enables us to resort to methods requiring continuous data input for our clusters recovery purposes. These methods are introduced in the next subsection.

\section{Clustering methods}\label{secclus}
We used the following clustering methods.

\subsection{Latent class analysis}
Latent class analysis operates directly on the presence-absence data. The method used here is the simplest of those that are referred to in the literature as ``latent class analysis'' \cite[chapter~3]{hagenaars}. It models the data as a mixture of locally independent Bernoulli distributions, i.e., the different variables/regions are assumed independent within clusters, although they can depend over the whole dataset. Data are modelled as i.i.d. according to the density \cite[chapter~6]{moustaki}:
\begin{align}\label{lcadens}
    f_{\mathbf{\eta}}(\mathbf{x}_i) = \sum_{k=1}^K \pi_k \prod_{j=1}^m \theta_{jk}^{x_{ij}}(1-\theta_{jk})^{1-x_{ij}},
\end{align}
where $\theta_{jk}$ is the probability of positive response for variable $j$ in cluster $k$, namely the probability that a species in the $k^{th}$ group inhabits the $j^{th}$ cell. The parameter vector ${\mathbf{\eta}}=((\pi_1,\theta_{11},\ldots,\theta_{1m}),\ldots,(\pi_K,\theta_{1K},\ldots,\theta_{mK}))$ stores mixture proportions and probabilities for presence per variable and cluster. The parameters can be fitted by Maximum Likelihood, using the EM-algorithm \cite{vermunt}. Observations can then be assigned to clusters by maximizing the estimated posterior probability of the observations to belong to the clusters \cite{hagenaars,moustaki}. The Bayesian Information Criterion (BIC) can be used to estimate the number of clusters, but in our study the number of clusters was taken as fixed and known. In our study we used the R-function \texttt{poLCA} in the R-package with the same name for computing this kind of latent class analysis. 

\subsection{Methods operating on distances}  
Some methods were used that take the Jaccard distance matrix as input, namely standard Single, Average, and Complete Linkage clustering \cite[Chapter~4]{everitt} with the dendrogram cut in such a way that the required number of clusters is produced. These were computed using the R-function \texttt{hclust}. Furthermore we used Partitioning Around Medoids (PAM; \cite{rousseeuw}) as computed by the R-function \texttt{pam}, and the K-modes algorithm \cite{huang97}, which tries to optimise the PAM objective function using the simple matching distance as computed by R-function \texttt{kmodes} in package \texttt{klaR} with parameter \texttt{fast} set to FALSE to avoid error messages.

\subsection{Methods operating on Euclidean data}
For clustering the Euclidean MDS-output, we used two of the most popular clustering methods, namely K-means \cite{jain} and the Gaussian mixture model \cite[chapter~6]{everitt}. K-means was computed by the R-function \texttt{kmeans}
(using parameters \texttt{nstart = 100, iter.max = 100} in order to allow for a more stable performance than granted by the default values), the Gaussian mixture model was fitted by the
function \texttt{Mclust} in package \texttt{mclust} \cite{mclust}. More precisely, data is assumed to be generated by a model with density
\begin{align}\label{gmmdens}
    f_{\mathbf{\eta}}(\mathbf{x}^{\texttt{mds}}_i)=\sum_{k=1}^K {\pi_k}\phi_{\mathbf{a}_k, \mathbf{\Sigma}_k}(\mathbf{x}^{\texttt{mds}}_i),
\end{align}
where $K$ is the number of mixture components, $\phi_{\mathbf{a}, \mathbf{\Sigma}}$ is the density of the multivariate Gaussian distribution with mean vector $\mathbf{a}$ and covariance matrix $\mathbf{\Sigma}$, $\pi_1,\ldots,\pi_K$ are the mixture proportions, and $\eta$ is a vector collecting all parameters. These parameters can be estimated by Maximum Likelihood as implemented in the EM-algorithm. \texttt{mclust} provides several constrained covariance matrix models (e.g., all covariance matrices equal) besides a fully flexible model, and the \texttt{mclust}- software uses the Bayesian Information Criterion (BIC) to select the best one \cite{mclust}. Observations are assigned to clusters by 
\begin{align}
    \hat{P}(c_i=k|\mathbf{x}^{\texttt{mds}}_i)=\frac{{\hat{\pi}_k}\phi_{\mathbf{\hat{a}}_k, \mathbf{\hat{\Sigma}}_k}(\mathbf{x}^{\texttt{mds}}_i)}{f_{\mathbf{\hat{\eta}}}(\mathbf{x}^{\texttt{mds}}_i)}
\end{align}
and estimating the cluster membership labels as follows:
\begin{align}
    \hat{c_i} = \argmax_{k\in\{1,\dots,K\}} \hat{P}(c_i=k|\mathbf{x}^{\texttt{mds}}_i).
\end{align}
The BIC is also often used to estimate the number of clusters, but this is treated as fixed and known here, which also means that there is no need to use one of the many available methods (that often lead to contradictory results) to decide the number of clusters for the other methods \cite{halkidi}.

Furthermore, \cite{AzzMen16} have proposed to run the density-based clustering method \texttt{pdfCluster} (R-function and package name) \cite{AzzTor07} on MDS-output in order to handle categorical or binary data. The method estimates the density by kernel methods. Clustering is then performed by finding density level sets using neighbourhood graphs. As opposed to the other methods in the study, pdfCluster will implicitly determine the number of clusters rather than allowing for it to be fixed. In order to avoid errors, we used the parameter settings \verb|graphtype="pairs"| and \verb|n.grid| equal to the number of observations.

K-means, the Gaussian mixture model, and pdfCluster are combined with the different MDS techniques and two dimensionalities of the MDS output. The combinations are called kc2, ks2, kc3, ks3, gc2, gs2, gc3, gs3, pdfc2, pdfs2, pdfc3, pdfs3 in the results section. ``k'', ``g'', ``pdf'' refer to K-means, the Gaussian mixture, and pdfCluster, respectively. ``c'' and ``s'' denote \texttt{cmsscale}, and \texttt{smacof}'s \texttt{mds}, respectively. The number in the end is the number of MDS dimensions used.    

\section{The simulation}
\label{sec3}
In order to compare the methods, we simulate datasets of $n$ 
species, each of which
is represented by a presence-absence vector of length $m$, and cluster them. 

Simulations enable us to investigate the features of the data that drive the performance of the clustering methods. 
Real biogeographical data sets can make this assessment difficult, as it is hard to separate results due to singularities in data from those due to methods properties \cite{casagranda}. Among the plethora of aspects that can be considered, in this project we examined the level of clusters' overlap, the cluster size, the width of the cluster specific areas and the number of clusters: in particular, we compared a situation with three clusters endowed with their specific geographic area with a situation where on top of these three groups a cluster of universal spreaders is added, as often exist in reality \cite{chen}. Such a cluster consists of species that are widespread across the whole map. These species are not informative regarding biotic elements, however they qualify as clusters in a data analytic sense, namely as a set of observations that behaves in a certain way. In the following, we will call the clusters made up of species that are \textit{not} universal spreaders ``proper'', because these species inhabit a specific group of cells and can signal the existence of a biotic element. 


The 24 parameter combinations for the simulated data are described below, after the presentation of the algorithm that was used to generate this kind of data.

\subsection{Data generation}
The algorithm implemented to simulate each species resorts repeatedly to a random draw from a distribution that chooses one category of the $m$ possible ones, each of which has its own probability of being picked. At each step, one category has to be picked, and therefore these probabilities sum to one. This distribution is called here ``categorical distribution''. Let $j=1, {\dots}, m$ be the categories. Each category has a probability $p_j$ of being picked and $\sum_{j=1}^{m} p_j = 1$. The probability mass function of our categorical distribution is:
\begin{equation}\label{catdistr}
f_\mathbf{p}(y) = \prod_{j=1}^{m} {p_j}^{[y=j]}    
\end{equation}
where $[y=j]$ is equal to 1 if $y=j$ and 0 otherwise.

The species in our data set inhabit the geographic units (cells) of a map. To each of the $m$ categories corresponds a cell: at every drawing step, a category is picked and its associated binary indicator is set to 1, meaning that the species is present in this geographic unit of the map. A value of 0 means that the cell has not been drawn and the species is absent.

Each cluster will have its specific area on the map, consisting of a set of cells whose probability to be inhabited by the species of the cluster will be larger than elsewhere. To simplify, there will be a unique probability inside these areas ($p_{in}$) and a unique probability outside of them ($p_{out}$), with $p_{in} > p_{out}$. Once the cluster specific areas have been determined, the probability vectors specific to each cluster can be produced. 
Species belonging to a given cluster will mainly inhabit the cluster specific area, while their occurrence outside of it will be rarer. The rareness of species outside of their specific cluster area will depend on the level of overlap $\omega$, see below.

We define the following notation:
\begin{enumerate}[(a)]
    \item observations $\mathbf{x}_i=(x_{i1},\ldots,x_{im})$ taking values in the product set $\{0,1\}^m$;
    \item cluster labels $k=1,{\dots},K$;
    \item species range $r_i=\sum_{j=1}^mx_{ij}$, for $i=1,\dots,n$; 
    \item vector of species cluster labels $\mathbf{c} = (c_1, \ldots, c_n)$;
    \item vector of cell cluster labels $\mathbf{d} = (d_1, \ldots, d_m)$. A cell is given the label $k$ when it belongs to the specific area of a cluster, i.e., the set of cells with elevated probability for cluster $k$. The probability to be inhabited by a species belonging to that cluster is $p_{in}$. Each cell can belong to at most one (exactly one in this study) cluster specific area.
\end{enumerate}
The following parameters of the simulation are fixed in advance:
\begin{itemize}
    \item the number of species $n$ and the number of cells $m$. $n$ was either 300 or 400, see below. $m$ was 60;
    \item the level of overlap $\omega=\frac{p_{out}}{p_{in}}$. Three levels of overlap are used, i.e. low ($\omega=0.05$), medium ($\omega=0.2$) and high ($\omega=0.4$);
    \item the number of clusters $K$. This was either 3  (with $n=300$), or 
4 (with $n=400$). There were always 3 proper clusters, and a cluster of universal spreaders was added or not; 
    \item the number of species belonging to each cluster $k$, i.e., $|\{i: c_i = k\}|$ for $k=1,\dots,K$. Two possibilities are investigated: one with all clusters having 100 observations and one with clusters of various sizes (a small clusters with 50 species, one with 100 and one with 150 species). If a group of universal spreaders is present, it consists of 100 observations;
    \item the number of cells constituting each cluster specific area on the map, i.e., $|\{j: d_j = k\}|$ for $k=1,\dots,K$. Two possibilities are investigated, one with equal-sized specific areas (all areas 20 cells) and one with heterogeneous specific areas (10, 20, and 30 cells). With unequal area sizes and unequal cluster sizes, cluster sizes were chosen in line with the area sizes, i.e., the larger the cluster specific area, the larger the cluster.
\end{itemize} 
The range $r_i$ of the $i^{th}$ species is instead randomly drawn. If the species belongs to a proper cluster, it is uniformly drawn between 1 and the size of the cluster specific area. If it is a universal spreader ($u.s.$), its range is at least as large as the widest cluster specific area on the map, namely
    \begin{align*}
    r_{u.s.} = \max_{k} |\{j: d_j = k\}|,
    \end{align*}
and the maximum is $m$. As a consequence, a universal spreader will always cover a region of the map that involves more cells than the habitat of any proper species. These choices have been made inspired by real datasets of this kind that we have seen; particularly, species ranges are highly variable.

Here is the algorithm for the $i^{th}$ species draw, provided that it is not a universal spreader:
\begin{enumerate}
    \item pick $r_i$ according to a discrete uniform distribution with domain $\{1,\dots, |\{j: c_i=d_j\}|\}$;
    \item set $\mathbf{x}^0_i = \mathbf{0}_m$, namely the observation is initialized as a vector of zeros of length $m$;
    \item for $t=0,\dots, r_i-1$:
        \begin{enumerate}
            \item compute $p^t_{in, c_i}$ and $p^t_{out, c_i}$ such that:
                \\
                \\
                \arraycolsep=1.4pt\def\arraystretch{2.2}
                $\left\{\begin{array}{ll}
                \frac{p^t_{out, c_i}}{p^t_{in, c_i}} = \omega \\
                |\{j:d_j=c_i \land x^t_{ij}=0 \}|\cdot{p^t_{in, c_i}} + |\{j:d_j \ne c_i \land x^t_{ij}=0\}|\cdot{p^t_{out, c_i}} = 1
                \end{array}\right.$
            \item specify the vector $\mathbf{p}^t_i$ according to the following rule:
                \\
                \\
                $p^t_{ij}=\left\{\begin{array}{ll}
                p^t_{in, c_i}, & \text { if } c_i=d_j \land x^t_{ij}=0 \\
                p^t_{out, c_i}, & \text { if } c_i \ne d_j \land x^t_{ij}=0 \\
                0 & \text{ otherwise }
                \end{array}\right.$
            \item draw j from a categorical distribution (eq. \ref{catdistr}) with probability vector $\mathbf{p}^t_i$;
            \item set $x^{t+1}_{ij}=1$.
        \end{enumerate}
    \item set the $i^{th}$ observation in the data set equal to $\mathbf{x}^{r_i}_i$.
\end{enumerate}
Note that the probability vector $\mathbf{p}^t_i$ is updated after each draw in such a way that every category can only be drawn once. Step 3 makes sure that $\omega$ (i.e., the chosen ratio between the probability to draw a cell without and within the cluster specific area) is respected by the probability vector $\mathbf{p}^t_i$. Species generated in this way may well have presences outside the cluster specific area, and absences within it, although the probability to be present within it is larger. This reflects realistically observed patterns.

The algorithm for the draw of the $i^{th}$ observation being a universal spreader is slightly different:
\begin{enumerate}
    \item set $\mathbf{x}^0_i = \mathbf{0}_m$, namely the observation is initialized as a vector of zeros of length $m$;
    \item pick $r_i$ according to a discrete uniform distribution with domain $\{r_{u.s.},\dots, m\}$;
    \item for $t=0,\dots,r_i-1$:
    \begin{enumerate}
        \item pick $j$ according to a discrete uniform distribution with domain $\{1,\dots, m\} \setminus \{j: x^t_{ij} = 1\}$;
        \item set $x^{t+1}_{ij}=1$.
    \end{enumerate}
    \item set the $i^{th}$ observation in the data set equal to $\mathbf{x}^{r_i}_i$.
\end{enumerate}

\subsection{Scenarios}
24 simulation scenarios have been used, which are listed in 
Table \ref{tab:24cases}.

\begin{table}[tb]
    \centering
    \caption{The 24 simulation scenarios resulting from the combination of the investigated data parameters.}
    \label{tab:24cases}
    \vspace{.1cm}
\begin{tabular}{r|c|c|c|c||r|c|c|c|c}
scenario & $\omega$ & $K$ & sizes & areas & case & $\omega$ & $K$ & sizes & areas\\
\hline\hline
1  & 0.05     & 3        & $=$   & $=$ & 13 & 0.05     & 3        & $=$   & $\neq$ \\
2  & 0.20  & 3        & $=$   & $=$ & 14 & 0.20  & 3        & $=$   & $\neq$ \\
3  & 0.40    & 3        & $=$   & $=$ & 15 & 0.40    & 3        & $=$   & $\neq$ \\
4  & 0.05     & 3+$u.s.$  & $=$   & $=$ & 16 & 0.05     & 3+$u.s.$  & $=$   & $\neq$ \\
5  & 0.20  & 3+$u.s.$  & $=$   & $=$ & 17 & 0.20  & 3+$u.s.$  & $=$   & $\neq$ \\
6  & 0.40    & 3+$u.s.$  & $=$   & $=$ & 18 & 0.40    & 3+$u.s.$  & $=$   & $\neq$ \\
7  & 0.05     & 3        & $\neq$ & $=$ & 19 & 0.05     & 3        & $\neq$ & $\neq$ \\
8  & 0.20  & 3        & $\neq$ & $=$ & 20 & 0.20  & 3        & $\neq$ & $\neq$\\
9  & 0.40    & 3        & $\neq$ & $=$ & 21 & 0.40    & 3        & $\neq$ & $\neq$ \\
10 & 0.05     & 3+$u.s.$  & $\neq$ & $=$ & 22 & 0.05     & 3+$u.s.$  & $\neq$ & $\neq$ \\
11 & 0.20  & 3+$u.s.$  & $\neq$ & $=$ & 23 & 0.20  & 3+$u.s.$  & $\neq$ & $\neq$ \\
12 & 0.40    & 3+$u.s.$  & $\neq$ & $=$ & 24 & 0.40    & 3+$u.s.$  & $\neq$ & $\neq$ \\
\hline
    \end{tabular}
\end{table}
Equal ($=$) group sizes imply that there are 100 observations per cluster (be it a proper one or a group of universal spreaders). When the group sizes are different ($\neq$), there is a cluster with 50 species, one with 100, and one with 150 (there are always 100 $u.s.$ species). When the areas are equal ($=$), the 60 cells are evenly divided across the clusters; when the areas differ ($\neq$), the smallest cluster specific area is made up of 10 cells/variables and the biggest one consists of 30 cells/variables. 
\begin{figure}[tb]
    \begin{subfigure}[t]{.5\linewidth}
  \centering
  \includegraphics[height=0.25\textheight,keepaspectratio]{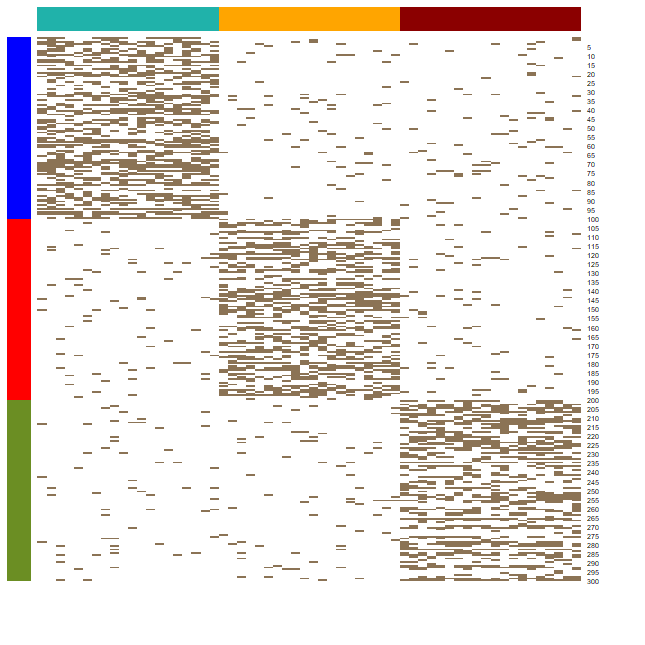}
  \captionsetup{width=.8\linewidth}
  \vspace{-5mm}
  \caption{Heatmap for case 1: no $u.s.$ group, equal sizes, equal areas, low overlap.}
   \label{fig:case1hm}
    \end{subfigure}
    \begin{subfigure}[t]{.5\linewidth}
  \centering
  \includegraphics[height=0.25\textheight,keepaspectratio]{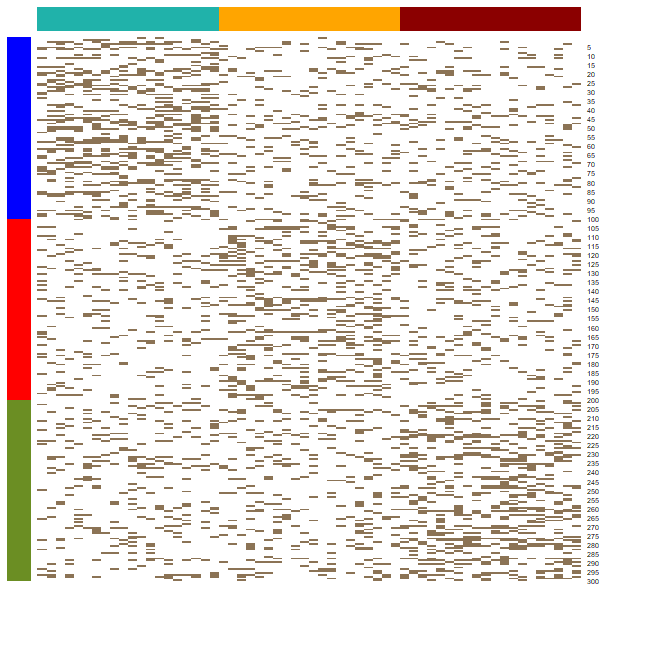}
  \captionsetup{width=.8\linewidth}
  \vspace{-5mm}
  \caption{Heatmap for case 3: no $u.s.$ group, equal sizes, equal areas, high overlap}
   \label{fig:case3hm}
    \end{subfigure}
    \vspace{5mm}
    \begin{subfigure}[t]{0.5\linewidth}
  \centering
  \includegraphics[height=0.25\textheight,keepaspectratio]{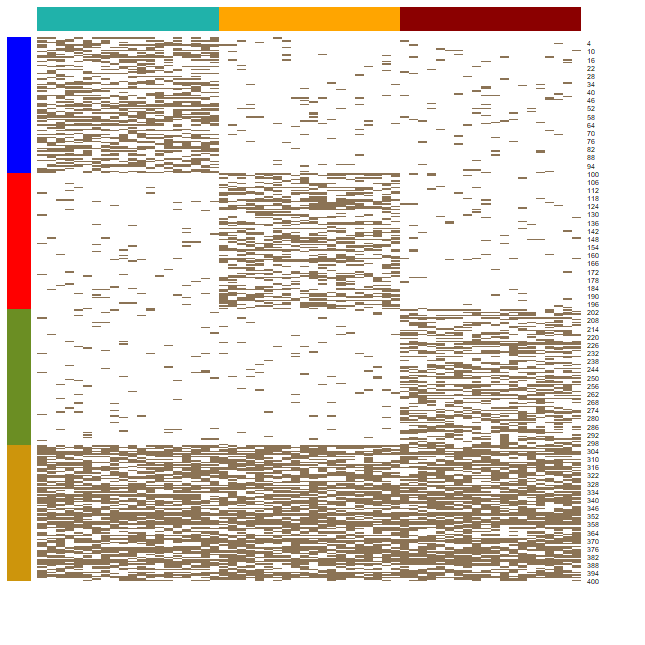}
  \captionsetup{width=.8\linewidth}
  \vspace{-5mm}
  \caption{Heatmap for case 4: with $u.s.$ group, equal sizes, equal areas, low overlap.}
   \label{fig:case4hm}
    \end{subfigure}
    \begin{subfigure}[t]{0.5\linewidth}
  \centering
  \includegraphics[height=0.25\textheight,keepaspectratio]{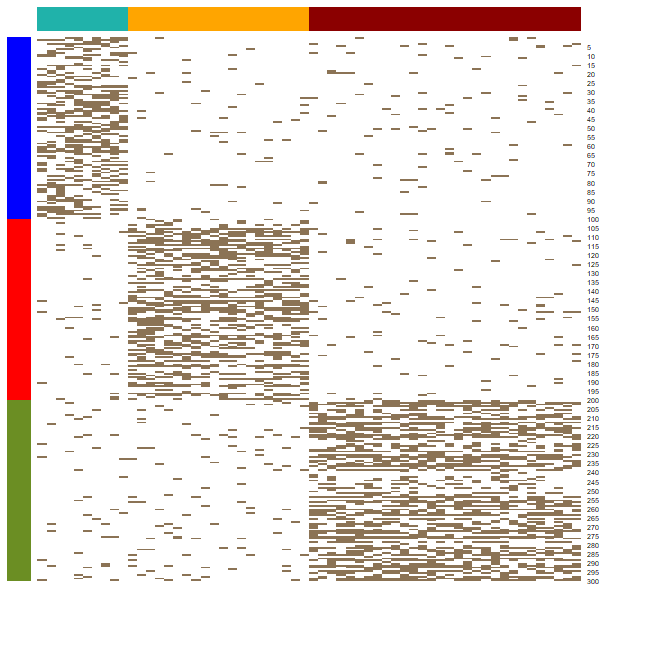}
  \captionsetup{width=.8\linewidth}
  \vspace{-5mm}
  \caption{Heatmap for case 13: no $u.s.$ group, equal sizes, different areas, low overlap.}
   \label{fig:case13hm}
    \end{subfigure}
    \captionsetup{width=.85\linewidth}
    \caption{Presence-absence heatmaps for cases 1, 3, 4 and 13. Rows correspond to species, columns correspond to cells. Colored bars represent species clusters (rows) and cluster specific areas (columns).}
    \label{fig:cs13413}
\end{figure}

We visualize the spread of the species in four exemplary cases using heatmaps. They represent the species in the rows and the cells in the columns: in each row, grey units represent cells where the species is present. 
In Figure \ref{fig:case1hm} the overlap is low and all groups have the same number of members and cluster specific cells. In Figure \ref{fig:case3hm} the high level of overlap makes it rather tough to distinguish the cluster specific areas.
Case 4 (Figure \ref{fig:case4hm}) differs from Case 1 in that it includes a group of universal spreaders. In Case 13 (Figure \ref{fig:case13hm}), clusters have specific areas of differing sizes, therefore species belonging to the green cluster will be able to inhabit up to thirty cells.

\section{Results}
\label{sec4}
\subsection{General results}
The simulations were evaluated using the Adjusted Rand Index \cite{hubert} in order to compare clusterings generated by the methods to the true clusterings, a measure of similarity between data clusterings that ranges from -1 to 1: two independent random partitions are expected to return an ARI of zero, whereas an ARI equal to 1 implies identical clusterings. Not only did we try to investigate what methods led to a satisfactory clustering recovery, but we also checked what data features had a significant impact on the methods' performance.

Each of the 24 scenarios reported in Table \ref{tab:24cases} was simulated ten times. This configuration generated a data set with 3120 rows: thirteen methods applied on 24 scenarios, each simulated ten times.

\begin{figure}[tb]
    \begin{subfigure}[t]{0.5\linewidth}
  \centering
  \includegraphics[height=0.30\textheight,keepaspectratio]{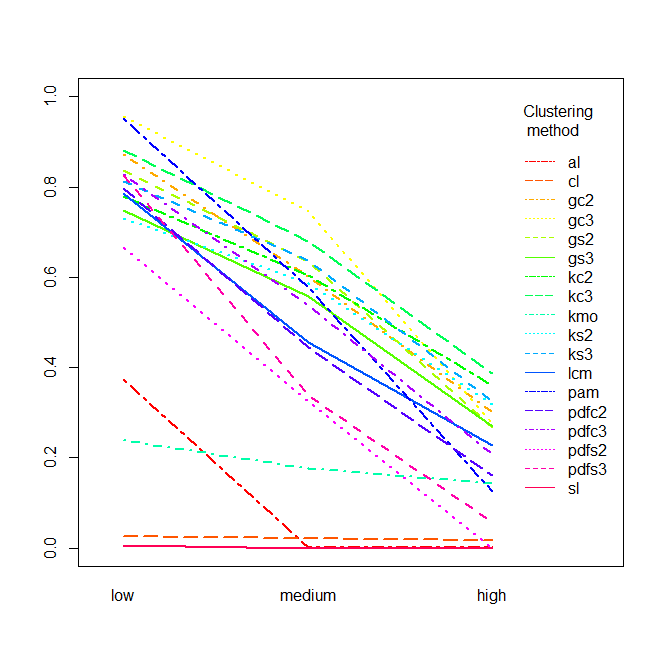}
  \captionsetup{width=.8\linewidth}
  \vspace{-5mm}
  \caption{Interaction plot showing the average ARI values (vertical axis) for the three \textbf{levels of overlap ($\omega$)}. }
   \label{fig:int_omega}
    \end{subfigure}
    \begin{subfigure}[t]{0.5\linewidth}
  \centering
  \includegraphics[height=0.30\textheight,keepaspectratio]{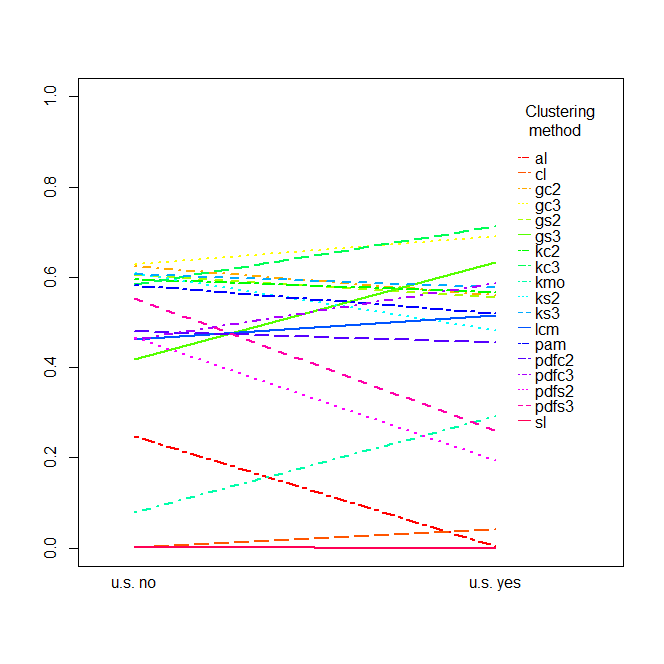}
  \captionsetup{width=.8\linewidth}
  \vspace{-5mm}
  \caption{Interaction plot showing the average ARI values (vertical axis) depending on the presence (yes) or absence (no) of \textbf{universal spreaders} in the simulation setup.}
   \label{fig:int_us}
    \end{subfigure}
    \begin{subfigure}[t]{0.5\linewidth}
  \centering
  \includegraphics[height=0.30\textheight,keepaspectratio]{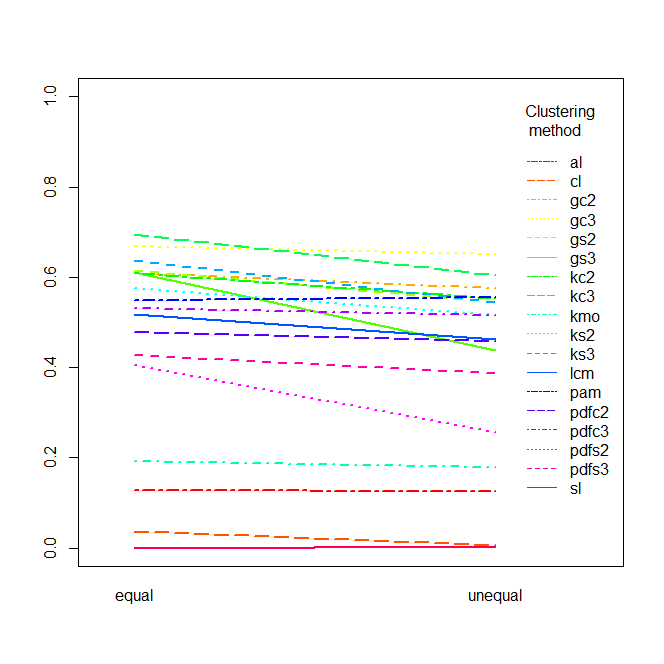}
  \captionsetup{width=.8\linewidth}
  \vspace{-5mm}
  \caption{Interaction plot showing the average ARI values (vertical axis) with equal or unequal \textbf{cluster sizes}.}
   \label{fig:int_sizes}
    \end{subfigure}
    \begin{subfigure}[t]{0.5\linewidth}
  \centering
  \includegraphics[height=0.30\textheight,keepaspectratio]{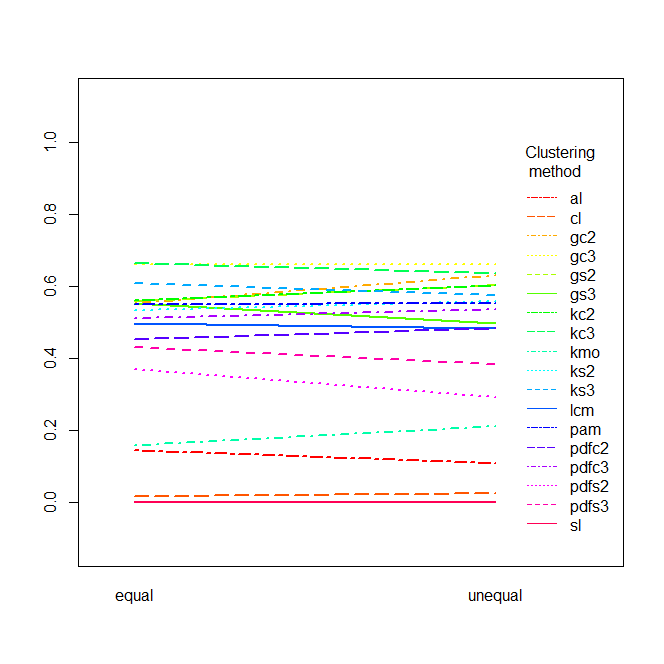}
  \captionsetup{width=.8\linewidth}
  \vspace{-5mm}
  \caption{Interaction plot showing the average ARI values (vertical axis) with equal or unequal \textbf{cluster specific area widths}.}
   \label{fig:int_areas}
    \end{subfigure}
    \captionsetup{width=.85\linewidth}
    \caption{Interaction plots of the average ARI values in the 24 cases (each simulated ten times). The lines referring to each of the twelve clustering recovery methods are indicated in the legends.}
    \label{fig:intplots}
\end{figure}
\begin{figure}[tb]
\makebox[\textwidth][c]{
    \centering
    \includegraphics[trim={0cm 1cm 1cm 1cm},clip,height=0.28\textheight,keepaspectratio]{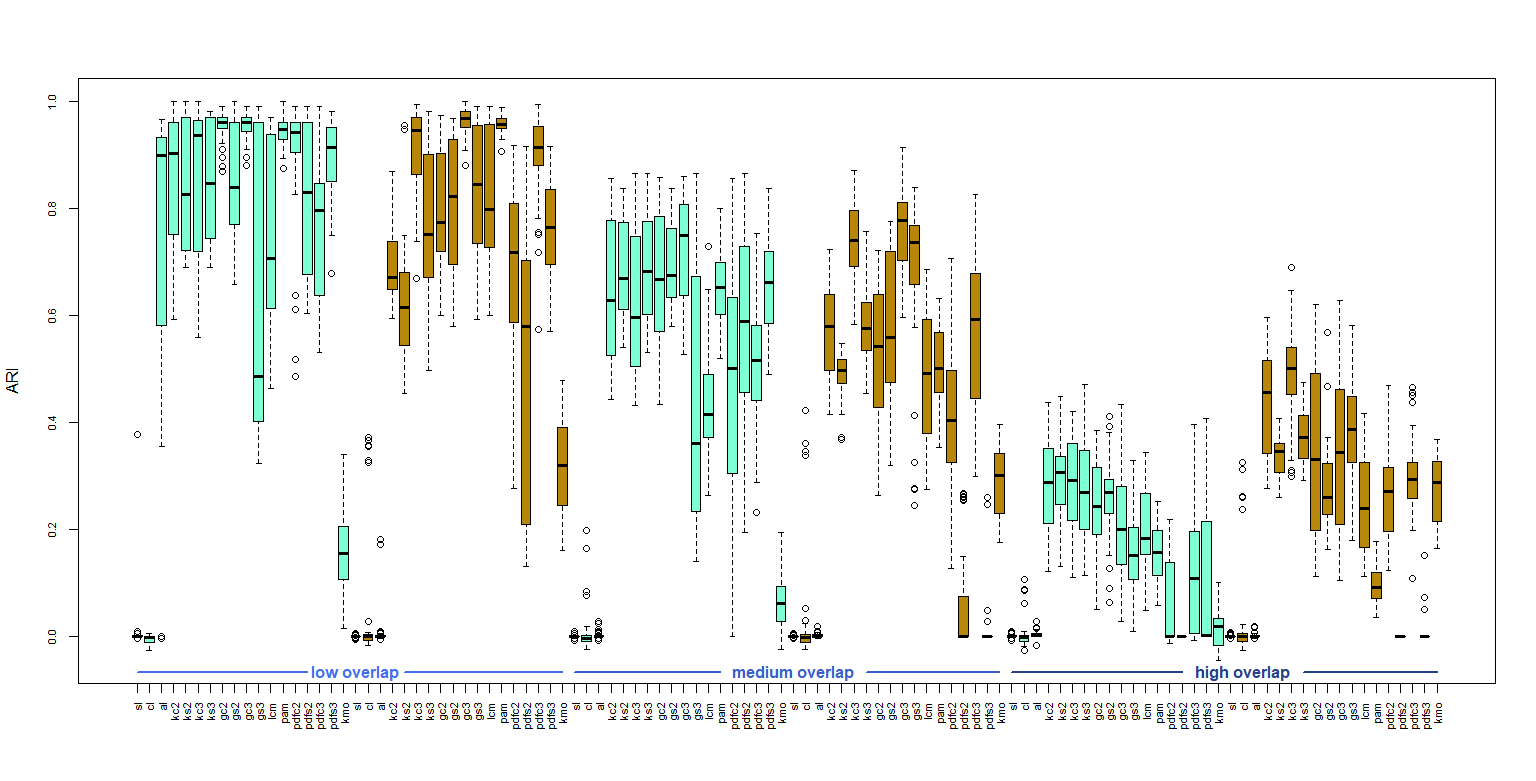}
    }
    \captionsetup{width=.85\linewidth}
    \caption{Boxplots of the ARIs achieved by the clustering methods, grouped by level of overlap and presence of a cluster of universal spreaders (sea-green boxplots: no $u.s.$ group; golden boxplots: with $u.s.$ group).}
    \label{fig:boxplots}
\end{figure}

Figure \ref{fig:intplots} shows interaction plots of the results. They show the ARI means on the vertical axis and the levels of the various data features on the horizontal axis. The different lines in the plots refer to the clustering methods. Figure \ref{fig:boxplots} shows boxplots of the ARI values achieved by the methods by overlap and presence of universal spreaders.

Unsurprisingly, methods are generally better with lower overlap, while some methods are helped whereas others are harmed by universal spreaders. Overall patterns regarding cluster sizes and cluster specific area widths are not that striking.

The performance of hierarchical clustering is generally rather bad, and 
clearly dominated by the other methods. Average linkage performs somewhat better than the other two methods for low overlap and without universal spreaders, but it is still worse than all the non-hierarchical methods; in fact the interaction plots could be interpreted as showing two clusters of methods, namely the hierarchical methods achieving low ARIs, and most other methods achieving better ARIs with K-modes and pdfs2 in between. We explore the reasons for this in Section \ref{shierarchical}.

K-modes performs a bit better than the hierarchical methods, but worse than the other methods. It becomes mostly better with universal spreaders. In particular, the use of the simple matching distance seems to be a worse choice for this kind of data than Jaccard for PAM.

The latent class analysis belongs to the cluster of better methods, but it
cannot compete with K-means and the Gaussian mixture after MDS, as can be seen in particular in Figure  \ref{fig:int_areas}. It does absolutely and relatively better with low overlap. PAM is among the best methods with low overlap, but deteriorates even stronger than the latent class analysis with more overlap. 

Out of the different variants of K-means and the Gaussian mixture model, regarding the overall average ARI those based on a 3-dimensional classical MDS (kc3, gc3) perform best, as can be seen in Figures \ref{fig:int_sizes}, \ref{fig:int_areas}. There is considerable variation over the different experimental factor levels though. Figure \ref{fig:boxplots} shows that gc2 is excellent with low overlap and without universal spreaders. kc3 performs much better with universal spreaders and does not stand out without them. Using the ratio (\texttt{smacof}) MDS is mostly worse than using classical MDS. Particularly it does not work well with the Gaussian mixture model and 3-dimensional MDS output with low and medium overlap and no universal spreaders, although universal spreaders help that method, particularly with medium overlap. Unequal sizes of cluster specific areas generally favour the classical MDS. The combination of medium overlap and universal spreaders generally favours the 3-dimensional MDS strongly, which otherwise has a small but not so substantial advantage. The K-means methods are overall competitive, but relatively weaker compared to the Gaussian mixture model for unequal cluster sizes. High overlap favours the K-means methods. pdfCluster is overall not quite as good as K-means and the Gaussian mixture. Overall it has a somewhat similar performance to the latent class analysis, with pdfc3 doing mostly better and the other three versions doing mostly worse. The impossibility to fix the number of clusters for pdfCluster puts it at a slight disadvantage, as the correct number of clusters was fixed for the other methods. pdfCluster mostly found the correct number of clusters, but occasionally (particularly with the \texttt{smacof}-MDS) it would put everything together in a single cluster. 

Figure \ref{fig:boxplots} gives some information about the variability of the results. pam and gc3 have most stable results for low overlap (without universal spreaders also gc2), whereas other methods, gs3 in particular, have substantial variation. Otherwise stability can mostly be found among bad results (e.g., the linkage methods).  
 
\subsection{More detailed insight}
\label{shierarchical}
\begin{figure}[tb]
  \centering
  \includegraphics[height=0.35\textheight,keepaspectratio]{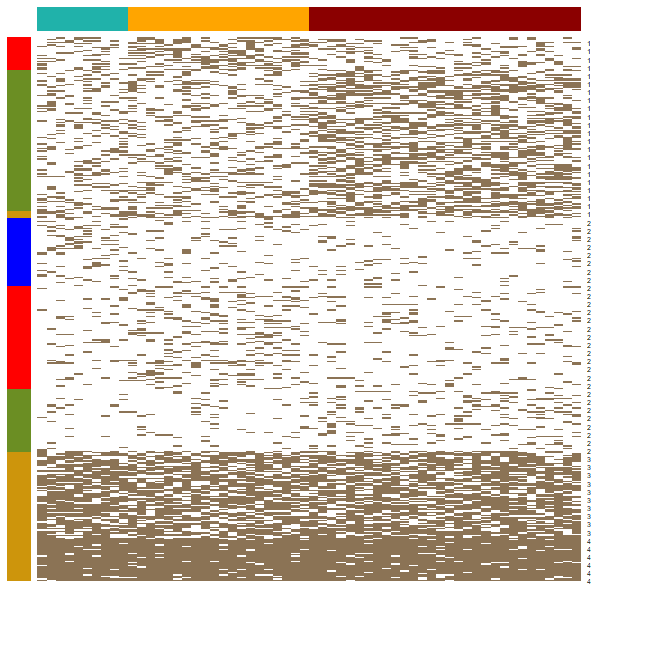}
  \vspace{-5mm}
    \captionsetup{width=.85\linewidth}
    \caption{Heatmap from a simulation of scenario 24: high overlap, with $u.s.$ group, various sizes and areas. ARI$=0.3538$. Rows sorted by LCA clustering. True cluster membership on the left (blue=1, red=2, green=3, gold=$u.s.$), cluster specific areas on top (light-blue=1, orange=2, dark red=3), LCA cluster membership on the right (numbers).}
    \label{fig:lcmcase24}
\end{figure}
In order to explore behavior of some methods in more detail and to understand their weaknesses, we use some exemplary visualizations.

For examining the performance of latent class analysis, we consider a simulation from scenario 24 of Table \ref{tab:24cases} with high overlap and different cluster sizes. The heatmap in Figure \ref{fig:lcmcase24} shows the four cluster latent class solution indirectly by the order of rows; colors on the side denote the true clusters, colors on top denote the cluster specific areas. A typical behavior of latent class analysis with large overlap was to build a cluster with sparse species, i.e., with small ranges, regardless to which original cluster they belonged. The smaller true clusters were hard to identify. Latent class analysis treats presences and absences symmetrically, which leads to a tendency to group sparse species together; also species with large range are easily grouped together, which is a good thing with a universal spreader cluster. Therefore its existence helps the latent class analysis to often achieve a larger ARI. It can also happen, however, that the universal spreaders cluster is split up into two clusters (relatively larger and smaller ranges), as happened  in Figure \ref{fig:lcmcase24}. The symmetric treatment of presences and absences is not good for identifying the proper clusters, at least not with large overlap. These rely on presences much more than on absences, and therefore the focus of the Jaccard distance on joint presences  rather than joint absences will help.

\begin{figure}[tb]
    \begin{subfigure}[t]{0.5\linewidth}
  \centering
  \includegraphics[trim={2cm 3cm 1cm 2cm},clip,height=0.25\textheight,keepaspectratio]{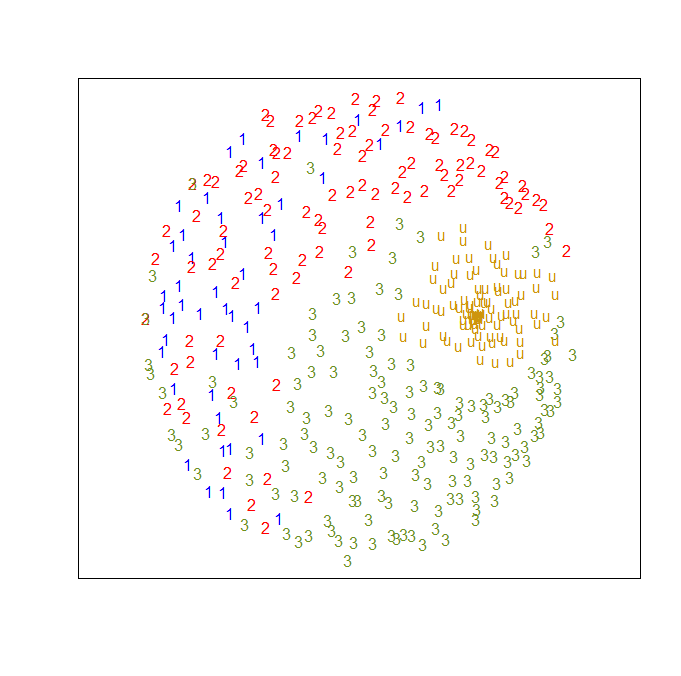}
  \captionsetup{width=.8\linewidth}
  \caption{Ratio \texttt{smacof} 2D mapping colored according to \textbf{true cluster membership}. The \texttt{u}'s are universal spreaders.}
   \label{fig:lcmsmatrue}
    \end{subfigure}
    \begin{subfigure}[t]{0.5\linewidth}
  \centering
  \includegraphics[trim={2cm 3cm 1cm 2cm},clip,height=0.25\textheight,keepaspectratio]{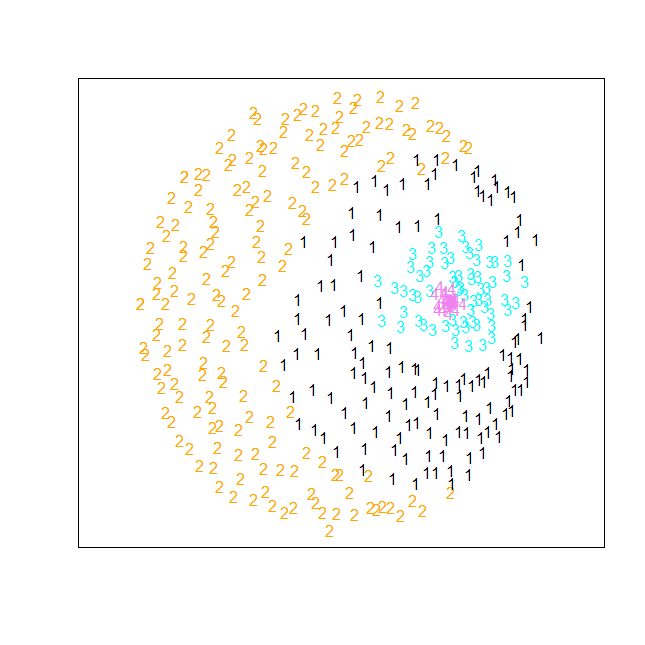}
  \captionsetup{width=.8\linewidth}
  \caption{Ratio \texttt{smacof} 2D mapping colored according \textbf{LCA clustering}.}
   \label{fig:lcmsmalcm}
    \end{subfigure}
    \begin{subfigure}[t]{0.5\linewidth}
  \centering
  \includegraphics[trim={2cm 3cm 1cm 2cm},clip,height=0.25\textheight,keepaspectratio]{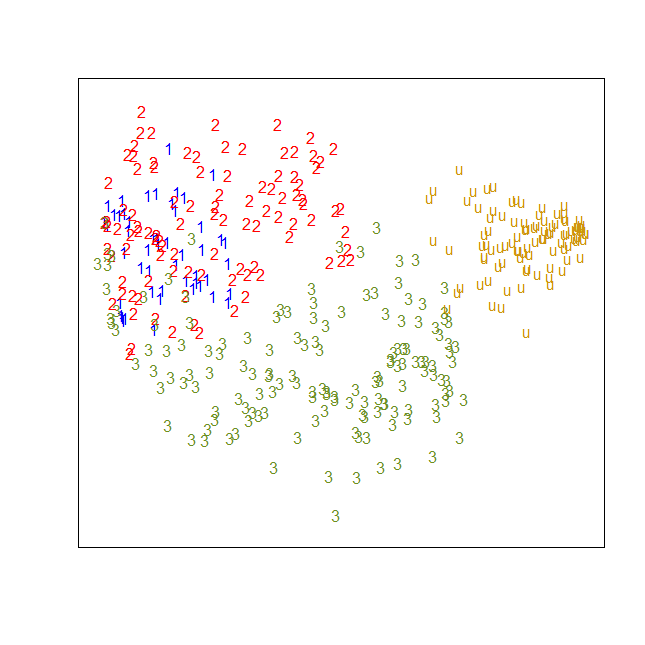}
  \captionsetup{width=.8\linewidth}
  \caption{Classical scaling 2D configuration colored according to \textbf{true cluster membership}. The \texttt{u}'s are universal spreaders.}
   \label{fig:lcmcmdtrue}
    \end{subfigure}
    \begin{subfigure}[t]{0.5\linewidth}
  \centering
  \includegraphics[trim={2cm 3cm 1cm 2cm},clip,height=0.25\textheight,keepaspectratio]{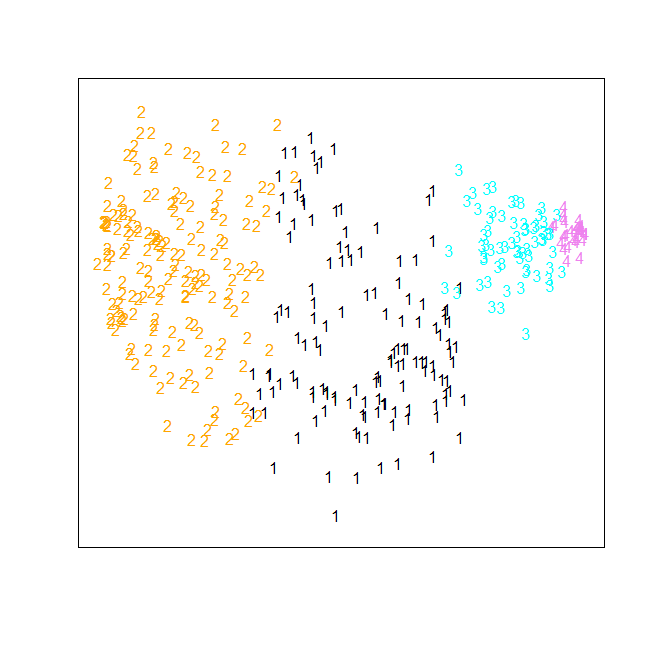}
  \captionsetup{width=.8\linewidth}
  \caption{Classical scaling 2D configuration colored according to \textbf{LCA clustering}.}
   \label{fig:lcmcmdlcm}
    \end{subfigure}
    \captionsetup{width=.85\linewidth}
    \caption{Ratio \texttt{smacof} MDS and classical scaling, based on the Jaccard's distance matrix computed on the data set simulated under scenario 24 (heatmap in Figure \ref{fig:lcmcase24}), colored by true clustering (left) and LCA clustering (right).}
    \label{fig:lcm_mds}
\end{figure}

Figure \ref{fig:lcm_mds} shows two-dimensional MDS maps of these data, highlighting the true cluster memberships ((a) and (c)) and the latent class clusters ((b) and (d)), for classical and ratio (\texttt{smacof}) MDS. Although not based on the MDS input, the latent class solution looks geometrically sensible in terms of both MDS visualizations, actually more so than the true clusters. However, latent class clusters 1 and 2 are very heterogeneous and involve large distances, merging observations from two or more true clusters together. This is in line with the fact that for large overlap with universal spreaders, starting from the MDS output, K-means, which prioritizes within-cluster homogeneity over separation and geometrically visible cluster shapes, does better than the Gaussian mixture model (Figure \ref{fig:boxplots}). The true clusters 1 and 2 are apparently hard to separate for any method, but they are still more homogeneous than latent class cluster 2, which holds the majority of both of them. 

Figure \ref{fig:lcm_mds} also serves for understanding why classical scaling performs better in our experiments. Geometrically, the ratio MDS is not worse at arranging the true clusters into a sensible shape, but both K-means and the Gaussian mixture model are connected to multivariate Gaussian distributions, which are characterised by linear relations between the variables, whereas the ratio MDS produces nonlinear cluster shapes and boundaries between true clusters. The ratio MDS in itself does not fail, but it does not serve as well as input for the Euclidean cluster methods. The classical MDS also separates the universal spreaders on one side of the plot, whereas in the ratio MDS they are, although also nicely separated, surrounded by observations from the proper clusters. Presence of universal spreaders helped some methods to achieve better results simply because they are often more easily ``visible'' as clusters to the methods.

\begin{figure}[tb]
\makebox[\textwidth][c]{
    \begin{subfigure}[t]{.5\linewidth}
  \centering
  \includegraphics[trim={0cm 1.5cm 0cm 1.5cm},clip,height=0.28\textheight,keepaspectratio]{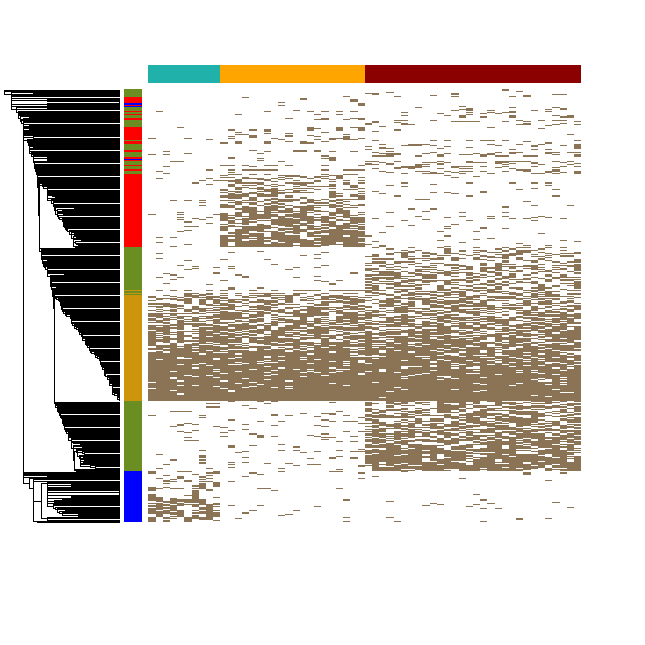}
  \captionsetup{width=.8\linewidth, justification=centering}
  \vspace{-10mm}
  \caption{Single linkage clustering. \\ ARI$=-0.0039$.}
   \label{fig:hc22heatsl}
    \end{subfigure}
    \medskip
    \begin{subfigure}[t]{.5\linewidth}
  \centering
  \includegraphics[trim={0cm 1.5cm 0cm 1.5cm},clip,height=0.28\textheight,keepaspectratio]{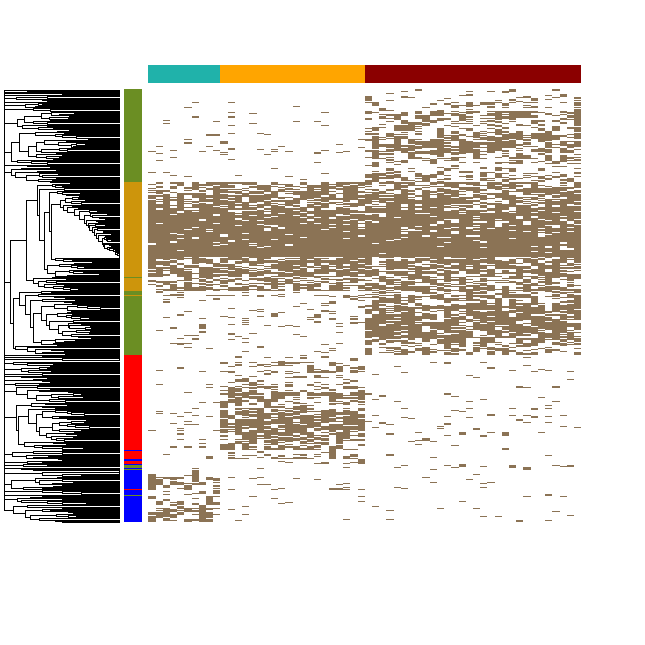}
  \captionsetup{width=.8\linewidth, justification=centering}
  \vspace{-10mm}
  \caption{Complete linkage clustering. \\ ARI$=-0.0051$.}
   \label{fig:hc22heatcl}
    \end{subfigure}
    }
    \begin{subfigure}[t]{\linewidth}
  \centering
  \includegraphics[trim={0cm 2cm 0cm 2cm},clip,height=0.28\textheight,keepaspectratio]{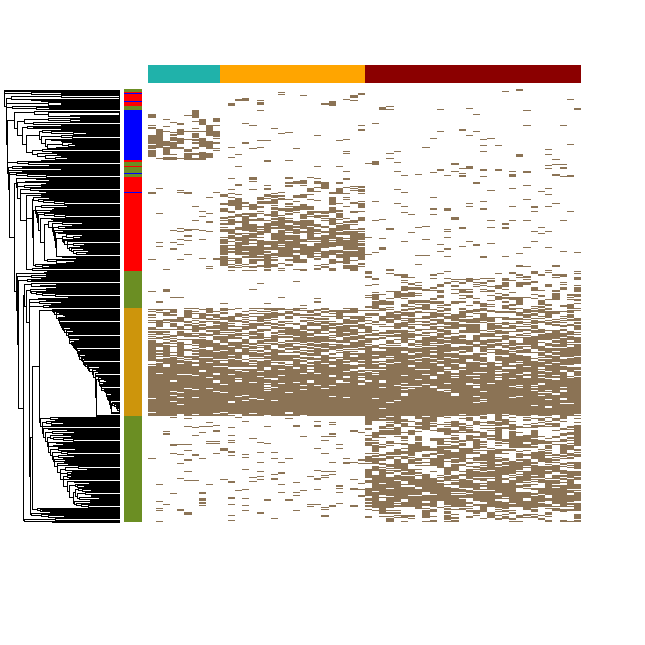}
  \captionsetup{width=.5\textwidth}
  \vspace{-5mm}
  \caption{Average linkage clustering. ARI$=0.0004$.}
   \label{fig:hc22heatal}
    \end{subfigure}
    \captionsetup{width=.85\linewidth}
    \caption{Heatmaps of a simulation of scenario 22 ($\omega=0.05$, with $u.s.$ cluster, various cluster sizes and areas), sorted according to the three hierarchical clustering solutions. Colours on the left side indicate true cluster membership of the species, while those on top refer to the cluster specific area of the cells.}
    \label{fig:hc22heatmaps}
\end{figure}

Regarding the weak performance of the hierarchical methods, Figure \ref{fig:hc22heatmaps} shows three heatmaps from data from scenario 22, with low overlap, and universal spreaders, and clusters have different sizes and different widths of their specific areas. These are shown together with the dendrograms form the three hierarchical clusterings. 

Single linkage (Figure \ref{fig:hc22heatsl}) clearly suffers from its well known
chaining issue \cite[Chapter~4]{everitt}. Fixing the number of clusters at 4, all observations are merged, and only a few very sparse species with a large Jaccard distance to all the other species are isolated as one-point clusters. Looking at the colors indicating the true clusters, the dendrogram as a whole is not perfect, but to some extent in line with the true clusters. A single clustering obtained by cutting the dendrogram at a certain height would however require a very large number of clusters to have at least the bigger ones of them in some agreement with the true clusters. The same phenomenon occurred with most other datasets, causing ARIs generally close to zero.

Complete linkage is not affected by chaining, but
Figure \ref{fig:hc22heatcl} shows a different problem. The Jaccard distance assigns its maximum value 1 to any two species that do not have joint presences. As occurs often in real data, the simulated data have several species with a small range that easily have no overlap. Complete linkage can't therefore join them in the same cluster, and they are still separated at the top level of the dendrogram. The dendrogram as a whole can be seen as even better in line with the true clustering, but in order to reflect this in a clustering with a fixed number of clusters, the method would need to integrate some sparse species with some clusters that include species to which they have the maximum Jaccard distance, and complete linkage cannot do that.

Unfortunately, average linkage (Figure \ref{fig:hc22heatal}) cannot solve both of these problems at the same time, but is rather also affected by chaining, isolating some sparse species when cutting for lower numbers of clusters. Once more, some useful agreement with the true clustering can be found in the dendrogram as a whole, but with a fixed number of clusters 4, the ARI is about zero, and 
a too large number of clusters would be needed to achieve a substantially better value. 

The best ARIs that could have been achieved by optimal cutting are 0.427 for single linkage (181 clusters), 0.469 for complete linkage (35 clusters), and 0.522 for average linkage (81 clusters). These are still lower than those for the MDS-based methods, latent class analysis, and PAM, largely due to the many small clusters involved. Still it shows that the results of the hierarchical methods are ultimately also connected to the true clusterings, though this requires looking deeper into the dendrogram.

A major lesson to learn from these insights regards the role of the Jaccard distance. On one hand, it is advantageous by treating presences and absences in an asymmetric manner. The latent class analysis suffers from treating them symmetrically, compared to the MDS-based methods. On the other hand, it also creates problems, particularly for the hierarchical methods, by assigning a maximum distance to pairs of sparse species that have no cell in common. \cite{haus03,hennig04} have preferred the Kulczynski dissimilarity, which is not a metric, but this also assigns maximum dissimilarity if there is no presence in common, although it may help by assessing sparse species as closer to species with a larger range of which they are a subset. \cite{hennig06} propose to involve geographical distances into the distance computation, and this may indeed improve matters, if only in a way that depends strongly on the geography of the specific real data, which will make setting up general models for simulation a more difficult task.

\section{Conclusions}
\label{sec5}
We have run a simulation study in order to evaluate the clustering performance, on multivariate presence-absence data, of a combination of multidimensional scaling and clustering methods for quantitative data (K-means, Gaussian mixture models, and pdfCluster), compared with methods that either operate directly on the data (such as the latent class analysis), or on a Jaccard or simple matching distance matrix computed from the data (hierarchical methods, PAM, and K-modes). Two different MDS methods and MDS outputs of dimensionality 2 and 3 were involved. 

The results suggest that the MDS-based techniques can be a valuable tool to cluster such data, returning by and large better ARI values than the competitors. The hierarchical methods did particularly badly, mainly due to the fact that cutting the dendrogram at the correct number of clusters proved inadequate; however it would be hard to repair this problem by any automatic method for deciding the number of clusters, because the required number of clusters for achieving a better ARI would be very large, and therefore potentially undesirable by researchers who want a simple and interpretable clustering solution. K-modes worked slightly better than the hierarchical methods.

Latent class analysis proved to be competitive when the data features were not demanding, but its performance deteriorated faster than that of MDS-based methods in more complicated setups. Similarly, PAM performed very well for low overlap but markedly worse, also in relative terms, with higher overlap.

For the MDS-based methods, classical MDS did a better job than \texttt{smacof}'s ratio MDS, due to the fact that classical MDS arranged the data in such a way that clusters could be separated more linearly, which makes the job of K-means and the Gaussian mixture model easier. However, without any linearity requirement, pdfCluster shared this behaviour. Cluster methods such as spectral clustering \cite{spectral} that do not involve linearity could also be tried with ratio MD. 3 dimensions worked overall slightly better than 2, but there is no reason to believe that increasing the dimensionality above 3 will improve results substantially. Comparing Gaussian mixture models and K-means, there is no clear winner. Both of these approaches worked well in some situations, and both worked mostly better than pdfCluster; the fact that the latter estimates the number of clusters automatically may have affected its results though.  

Regarding characteristics of the data, unsurprisingly, increasing levels of overlap among clusters turned out to be the most important factor, making separation of the true clusters more difficult for all methods.

Interesting extensions for future research could be involving other distance measures and clustering methods, with possible alternatives suggested above, and in particular involving methods to estimate the number of clusters, although this is a hard task due to the large number of possible combinations between clustering methods for fixed $K$ and methods to choose $K$.

\end{document}